\documentclass[conference,letterpaper]{IEEEtran}

\IEEEoverridecommandlockouts
\usepackage[dvipsnames]{xcolor}
\usepackage[cmex10]{amsmath}
\usepackage{amssymb,amsfonts,amsthm}
\usepackage{mathtools}
\usepackage{relsize}
\usepackage{algorithmic}
\usepackage{url}
\usepackage{hyperref}
\usepackage{graphicx}
\usepackage{textcomp}
\usepackage{tikz}
\usepackage{subfig}
\usepackage{stfloats}
\usepackage{bbm}
\usepackage[inline]{enumitem}
\usepackage[flushleft]{threeparttable}
\usepackage{multirow}
\usepackage{diagbox}
\usepackage{float}
\newtheorem{prop}{Proposition}

\DeclareMathOperator*{\argmax}{arg\,max}
\DeclareMathOperator*{\argmin}{arg\,min}
\DeclarePairedDelimiter\abs{\lvert}{\rvert}%

\DeclarePairedDelimiterX{\infdivx}[2]{(}{)}{%
  #1\;\delimsize\|\;#2%
}
\newcommand{\infdiv}{KL\infdivx}

\makeatletter
\let\oldabs\abs
\def\abs{\@ifstar{\oldabs}{\oldabs*}}

\makeatother
\newcommand*\xor{\mathbin{\oplus}}

\usepackage{cite}

\begin{document}

\title{Data-Driven  Ensembles for Deep and Hard-Decision Hybrid Decoding}

\author{%
  \IEEEauthorblockN{Tomer~Raviv, Nir~Raviv, and~Yair~Be'ery,~\IEEEmembership{Senior~Member,~IEEE}}
  \IEEEauthorblockA{School of Electrical Engineering, Tel-Aviv University\\
                    Tel-Aviv 6997801, Israel\\
                    Email: \{tomerraviv95, nirraviv89\}@gmail.com, 
                    ybeery@eng.tau.ac.il}
}


\maketitle

\begin{abstract}
Ensemble models are widely used to solve complex tasks by their decomposition into multiple simpler tasks, each one solved locally by a single member of the ensemble. Decoding of error-correction codes is a hard problem due to the curse of dimensionality, leading one to consider ensembles-of-decoders as a possible solution. Nonetheless, one must take complexity into account, especially in decoding. We suggest a low-complexity scheme where a single member participates in the decoding of each word. First, the distribution of feasible words is partitioned into non-overlapping regions. Thereafter, specialized experts are formed by independently training each member on a single region. A classical hard-decision decoder (HDD) is employed to map every word to a single expert in an injective manner. FER gains of up to 0.4dB at the waterfall region, and of 1.25dB at the error floor region are achieved for two BCH(63,36) and (63,45) codes with cycle-reduced parity-check matrices, compared to the previous best result of \cite{be2019active}. 
\end{abstract}


\section{Introduction}

\begin{figure*}[hb]
\centering
    \includegraphics[width=376px,height=142px]{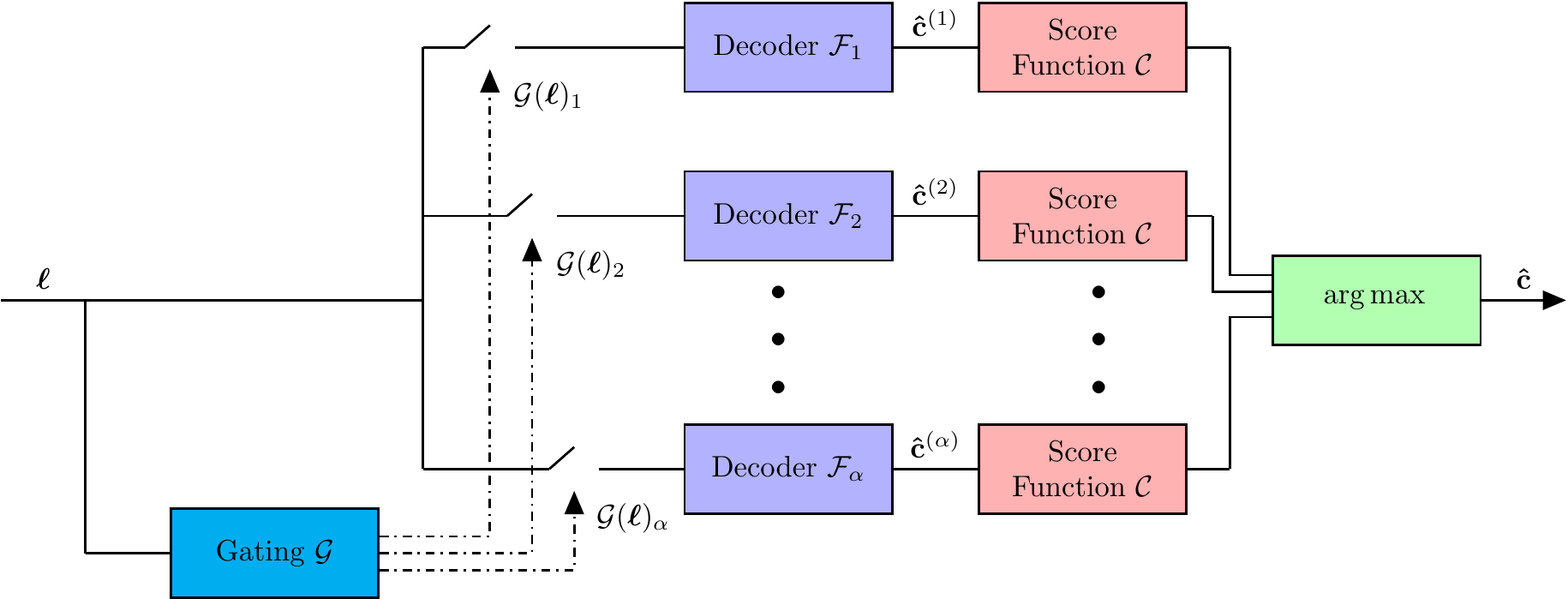}
    \caption{Data-Driven Ensemble of Decoders}
    \label{fig:ensemble_drawing}
\end{figure*}

Data-driven applications are essential in scenarios where mathematical exact models are unknown, or are too complicated to be manually derived \cite{simeone2018very,zappone2019wireless}. For instance, a Neural Network (NN) is used in \cite{belghazi2018mine} to estimate a tight lower bound on the mutual information between two high-dimensional continuous variables. Nonetheless, even where theoretical models are known, machine-learning models can be integrated  with classical algorithms in order to mitigate non-linear phenomena and discover high-dimensional and complex patterns. This is where the space of \textit{domain knowledge} and the space of \textit{machine-learning} intersect. 

Lately, active learning \cite{be2019active} was adopted to the Weighted Belief Propagation (WBP) algorithm \cite{nachmani2016learning,nachmani2018deep}. The method offered a trade off: higher decoding gains at inference for the cost of longer preprocess time solely in training. The intuition for the additional gains is that not all samples are equally important for the training of the decoder. The selection of training data is highly based on domain knowledge, as an expert-guided solution. This guidance differs the proposed methods from generic active learning methods that lack specific domain guidance. 

Accordingly, selection of data may prove useful not merely to a single WBP-model, but to an ensemble of them. The main intuition behind this divide-and-conquer approach is that combination of multiple \textit{diverse} members is expected to perform better than all individual basic algorithms that compose the ensemble. See Figure \ref{fig:ensemble_drawing} for an ensemble of decoders, whose architecture will be detailed in the next section. This paper addresses several issues with regard to ensembles:

\begin{itemize}

\item{How to encourage diversity among all members? Our approach in Section \ref{sec:Data Partitions Motivated By Domain Knowledge} suggests a \textit{partition of the data distribution}, hereafter each member specializes on a different part of the distribution.}

\item{To which specialized decoder should a received word be mapped? Section \ref{sec:GatingFunction} introduces the concept of \textit{gating}, discussing how to map any received channel word to the most-fitting decoder. Exploiting domain knowledge with an HDD is empirically seen as a reliable prior.}
\end{itemize}
At last, we demonstrate the performance of the proposed methods in Section \ref{sec:Results}.

\section{Background}

\subsection{Motivation}
We outline the motivation for ensemble-decoding by presenting main differences from a more familiar topic - the list decoding. For example, consider the belief propagation list (BPL) decoder for polar codes by Elkelesh et al. \cite{elkelesh2018belief}. All decoders in the BPL run in parallel since \textit{''there exists no clear evidence on which graph permutation performs best for a given input''} as the authors indicate. Were the decoders input-specialized, one could simply map each word to a single decoder, preserving computation resources. 

Furthermore, Arli and Gazi  \cite{arli2019noise} suggest adding stochastic perturbations with varying magnitudes to the received channel word, followed by applying the same Belief Propagation (BP) algorithm on each of the multiple copies. In fact, each BP decoder is introduced with a \textit{modified input distribution}. Ambiguity arises with respect to the optimal choices for the magnitudes of the artificial noises. One would want each decoder to correctly decode a different part of the original input distribution, such that the list-decoder covers the entire input distribution in an efficient manner.

\subsection{Notation}
We denote the $i^{\mathrm{th}}$ element of a vector $\mathbf{v}$ with a subscript $v_i$. Further, $v_{i,j}$ corresponds to an element of a matrix. However, denoted with a superscript, $\mathbf{v}^{(i)}$ presents the $i^{\mathrm{th}}$ member of a set.

We present all elements of a classical encoder-decoder network. Let $\mathbf{u} \in \{0,1\}^k$ be a message word encoded with function $\mathcal{U}:\{0,1\}^k\rightarrow\{0,1\}^V$ to form codeword $\mathbf{c}$, with $k$ and $V$ being the information word's length and the codeword's length, respectively. Denote by $\mathbf{x}$ the BPSK-modulated ($0\rightarrow 1,1\rightarrow -1$) transmitted word. After transmission through the AWGN channel the received word is $\mathbf{y} = \mathbf{x}+\mathbf{n}$, where $\mathbf{n}\sim N(\mathbf{0},\,\sigma^{2}\mathbf{I})$ is the white noise. Next, log-likelihood ratio (LLR) values are considered for decoding by $\boldsymbol{\ell} = \frac{2}{\sigma^2} \cdot \mathbf{y}$. At last, a decoding function $\mathcal{F}:\mathbb{R}^V\rightarrow\{0,1\}^V$ is applied to the LLR values to form the decoded codeword $\mathbf{\hat{c}} = \mathcal{F}(\boldsymbol{\ell})$. Also, one usually employs a stopping criterion after each decoding iteration.

We hereby introduce the components of data-driven ensembles for decoding. All relevant components are depicted in Figure \ref{fig:ensemble_drawing}. 
Consider a decoder $\mathcal{F}$ parameterized by weights $\boldsymbol{w}$, obtained by training $\mathcal{F}$ over dataset $\mathcal{D}$ until convergence. Notate this decoder $\mathcal{F}^{\boldsymbol{w},\mathcal{D}}$. Now, let $\{\mathcal{F}_1^{\boldsymbol{w}^{(1)},\mathcal{D}^{(1)}},\ldots,\mathcal{F}_\alpha^{\boldsymbol{w}^{(\alpha)},\mathcal{D}^{(\alpha)}}\}$ be a set of $\alpha$ weighted decoders, each decoder trained on a different dataset hence obtaining different parameters. Each decoder $\mathcal{F}_i^{\boldsymbol{w}^{(i)},\mathcal{D}^{(i)}}$is notated $\mathcal{F}_i$ in short. A word entering the ensemble is first mapped by a mapping function $\mathcal{G}:\mathbb{R}^V\rightarrow{\{0,1\}}^{\alpha}$. Next, decoder $i$ tries to decode $\boldsymbol{\ell}$ if $\mathcal{G}(\boldsymbol{\ell})_i=1$. As such, $\mathcal{G}$ is referred to as the \textit{gating function}. After decoding, a \textit{score function} $\mathcal{C}:\{0,1\}^V \rightarrow \mathbb{R}$ is employed, used to rank each of the decoded words. The decoded word maximizing this score among all valid candidates (a valid candidate is a codeword) is chosen as the final decoded word. If no valid candidates exist, all candidates are considered. The output of the ensemble can be explicitly written as:
\begin{equation} \label{eq:ensemble}
    \mathbf{\hat{c}} = \argmax_{\mathbf{\hat{c}}^{(i)},i \in \{j:\mathcal{G}(\boldsymbol{\ell})_j=1\}} {\mathcal{C}(\mathbf{\hat{c}}^{(i)})} 
\end{equation}
\begin{equation*}
    \mathbf{\hat{c}}^{(i)} = \mathcal{F}_i(\boldsymbol{\ell})
\end{equation*}
We choose $\mathcal{G}$ such that for every $\boldsymbol{\ell}$, at least one decoder is active. 

\subsection{WBP}

The BP \cite{pearl2014probabilistic} is an inference algorithm used to decode corrupted codewords in an iterative manner. The algorithm passes messages over the nodes of the Tanner graph until convergence or a maximum number of iterations is reached. The nodes in the Tanner graph are of two types: variable and check nodes. An edge exists between a variable node $v$ and a check node $h$ iff variable $v$ participates in the condition defined by the $h^{\mathrm{th}}$ row in the parity check matrix $\mathbf{H}$. Nachmani et al. \cite{nachmani2018deep,nachmani2016learning} assigned learnable weights to the BP algorithm. This formulation unfolds the BP algorithm into an NN, referred to as WBP. For a comprehensive explanation of the subject, please refer to \cite{nachmani2018deep,nachmani2016learning}.

\section{Ensembles}

Ensembles composed of weighted BP decoders allow for input-specialized experts. An optimal ensemble of decoders covers as much of the input distribution as possible, while minimizing the number of required decoders. We formulate the ensemble of decoders framework, presenting main components:

\begin{enumerate}[label=(\Alph*)]
    \item{\textit{Data partitions motivated by domain knowledge} - how to exploit domain knowledge in order to partition the input distribution into similar-context distributions.}
    \item{\textit{Gating function} - the choice of a function mapping from a received word to a distinct decoder.}
    \item{\textit{Combiner mechanism} - the ranking of candidate decoded codewords.}
\end{enumerate}

We highly recommend the book \cite{rokach2010pattern} and survey \cite{rokach2010ensemble} for a tutorial on ensembles in machine-learning.

\subsection{Data Partitions Motivated By Domain Knowledge}
\label{sec:Data Partitions Motivated By Domain Knowledge}

The diversity of the ensemble refers to the notion that each classifier specializes on a specific region of the data distribution. Rokach \cite{rokach2010pattern} indicated that diversity may be obtained by altering the presentation of the input space (distribution). 

Consider the distribution $P(\mathbf{e})$ of binary errors ${\mathbf{e}=\mathbf{y}_{\mathrm{HD}}\xor{\mathbf{c}}}$ at the channel's output, where $\mathbf{y}_{\mathrm{HD}}$ is the received word after a hard-decision rule ($\mathbb{R}^+\rightarrow 0,\mathbb{R}^-\rightarrow 1$). Denote by $\mathcal{E}=\{\mathbf{e}^{(1)},\ldots,\mathbf{e}^{(K)}\}$ the set of $K$ observable binary error patterns, only used in training. We seek a partition of $\mathcal{E}$ into $\alpha$ different error-regions: 
\begin{equation*}
    \mathcal{E}=\bigcup_{i=1}^{\alpha}\mathcal{X}^{(i)} : \mathcal{X}^{(i)}\cap\mathcal{X}^{(j)}=\emptyset, \forall{i\neq j}.
\end{equation*}
These regions induce $\alpha$ datasets $\{\mathcal{D}^{(1)},\ldots,\mathcal{D}^{(\alpha)}\}$ --- one training dataset per decoder --- according to the next relation:
\begin{equation}
\label{eq:datasets_induction}
    \mathcal{D}^{(i)} = \{\boldsymbol{\ell}^{(\kappa)}:\mathbf{e}^{(\kappa)}\in \mathcal{X}^{(i)}\}.
\end{equation}
The choice of the above partition is crucial not only to the performance of the single decoder, but to the generative capabilities of the overall ensemble. The rest of this section is devoted to the proposal of two different partitions.
\subsubsection{Hamming Distance Partition}
\label{subsubsec:HammingDistancePartition}

The Hamming distance $d_H$ is a widely-known metric encapsulating important knowledge - the number of bits positions differed between the hard-decision of the received word and the correct word. One simple approach is to partition the errors by the Hamming distance from the zero-errors vector:
\begin{equation*}
    \mathcal{X}^{(i)} = \{\mathbf{e}^{(\kappa)}:\mathbf{e}^{(\kappa)}\text{ has }i\text{ non-zero bits}\}.
\end{equation*}
with the datasets induced as in equation (\ref{eq:datasets_induction}). Furthermore, all patterns $\mathbf{e}^{(\kappa)}$ with more than $\alpha$ non-zero bits are assigned to $\mathcal{X}^{(\alpha)}$.

\subsubsection{Syndrome-Guided EM Partition}
\label{subsubsec:SyndromeGuidedEM}

Though the above partition is straightforward, we argue it is also too restrictive. The number of errors in a word is merely a single feature; one should consider all latent features responsible for successful decoding. This, unfortunately, is analytically infeasible.

Instead, we suggest to cluster together similar error patterns with the expectation-maximization (EM) \cite{bishop2006pattern,koller2009probabilistic} algorithm. Each cluster defines an error-region $\mathcal{X}^{(i)}$. Then, every decoder specializes in decoding words with similar errors: ones that belong to the same cluster. We elaborate on this below.

Let $\boldsymbol{\mu}^{(i)} \in [0,1]^V$ be a multivariate Bernoulli distribution corresponding to region $\mathcal{X}^{(i)}$. Let $\mathcal{R}=\{(\boldsymbol{\mu}^{(1)},\pi_1),\ldots,(\boldsymbol{\mu}^{(\alpha)},\pi_\alpha)\}$ be a Bernoulli mixture with $\pi_i \in [0,1]$ being each mixture's coefficient such that $\sum_{i=1}^\alpha \pi_i = 1$. We assume that each error $\mathbf{e}$ is distributed by mixture $\mathcal{R}$:
\begin{equation*}
    P(\mathbf{e}|\mathcal{R}) = \sum_{i=1}^{\alpha}{\pi_i P(\mathbf{e}|\boldsymbol{\mu}^{(i)})}
\end{equation*}
where the Bernoulli prior is:
\begin{equation*}
    P(\mathbf{e}|\boldsymbol{\mu}^{(i)}) = \prod_{v=1}^{V}{(\mu_v^{(i)})^{e_v}(1-\mu_v^{(i)})^{1-e_v}}.
\end{equation*}

At first, all $\boldsymbol{\mu}^{(i)}$ and $\boldsymbol{\pi}$ are randomly initialized. Then, the EM algorithm is applied to infer parameters that maximize the log-likelihood function over $K$ samples:
\begin{equation}
\label{eq:log_likelihood}
    \log{(\mathcal{E}|\mathcal{R})} = \\
    \sum_{\kappa=1}^K \log\left(P(\mathbf{e}^{(\kappa)}|\mathcal{R})\right).
\end{equation}
See Bishop \cite{bishop2006pattern} for more details on the EM algorithm.

This clustering is performed once as a preprocess phase. Upon convergence to some final parameters, region $\mathcal{X}^{(i)}$ is assigned with patterns more probable to be originated from cluster $i$ than from any other cluster $j$: 
\begin{equation*}
    \mathcal{X}^{(i)} = \{\mathbf{e}^{(\kappa)}:\pi_i P(\mathbf{e}^{(\kappa)}|\boldsymbol{\mu}^{(i)})>\pi_j P(\mathbf{e}^{(\kappa)}|\boldsymbol{\mu}^{(j)}), \forall{j\neq i}\}. 
\end{equation*}
Thereafter, each $\mathcal{D}^{(i)}$ is formed following equation (\ref{eq:datasets_induction}).

\begin{prop} \label{prop:naive_em}
Let $\mathcal{E}$ be formed of error patterns drawn from $\alpha$ different AWGN channels $\sigma^{(1)},\ldots,\sigma^{(\alpha)}$. Let $K$ be the number of total patterns, where an equal number is drawn from each channel. Then, for $\alpha$ desired mixture centers and as $K$ tends to infinity, the global maximum of the likelihood is attained at parameters ${\boldsymbol{\mu}^{(i)}=\Big(Q(\frac{1}{\sigma^{(i)}}),\ldots,Q(\frac{1}{\sigma^{(i)}})\Big)}$, $Q(\cdot)$ being the $Q$-function.
\end{prop}
\noindent{\textbf{Proof}. We shortly sketch the proof; see Appendix \ref{app:proof} for further details. First, the \textit{true} centers of the mixture were derived, recalling that the AWGN channel may be viewed as a binary symmetric channel with a crossover probability of $Q(\frac{1}{\sigma^{(i)}})$. Second, the parameterized centers were shown to attain the global maximum of the likelihood function when identical to the true centers, similarly to the analysis in \cite{srebro2007there}.}

Proposition \ref{prop:naive_em} indicates that though one is inclined to model the distribution of binary errors at the channel's output with a mixture of multivariate Bernoulli distribution, a naive application of the EM tends to converge to a trivial solution. \textit{This trivial solution fails to adequately cluster complex classes}.

We suggest using the code structure, which can be seen as \textit{a priori} knowledge, to find non-trivial latent classes. For each error, we first calculate the syndrome, $\mathbf{s} = \mathbf{H}\mathbf{e}$. Thereafter, each index $v$ is given a label in $\{0,1\}$ based on the majority of either unsatisfied or satisfied conditions it is connected to:
\begin{equation}
    q_v = \argmax_{b\in\{0,1\}}{\sum_{i \in \mathcal{N}(v)}\mathbbm{1}_{s_i=b}}
\end{equation}
with $\mathcal{N}(v)$ being the indices of check nodes connected to $v$ in the Tanner graph and $\mathbbm{1}$ denotes the indicator function (which has value 1 if $s_i=b$ and 0 otherwise). 

Assume each latent class $i$, which corresponds to a single error-region, is modeled with \textit{two different multivariate Bernoulli distributions} $\boldsymbol{\mu}^{(i,0)},\boldsymbol{\mu}^{(i,1)}$. Label $q_v$ determines for each index $v$ it's Bernoulli parameter $\mu_v^{(i,q_v)}$. Under this new model, the Bernoulli mixture $\mathcal{R}^{\mathrm{syn}}$ is:
\begin{equation*}
   \mathcal{R}^{\mathrm{syn}}=\{(\boldsymbol{\mu}^{(1,0)},\boldsymbol{\mu}^{(1,1)},\pi_1),\ldots,(\boldsymbol{\mu}^{(\alpha,0)},\boldsymbol{\mu}^{(\alpha,1)},\pi_\alpha)\}
\end{equation*}
having $\alpha$ latent classes:
\begin{equation*}
    P(\mathbf{e}|\mathcal{R}^{\mathrm{syn}}) = \sum_{i=1}^{\alpha}{\pi_i P(\mathbf{e}|\boldsymbol{\phi}^{(i)})}
\end{equation*}
where:
\begin{equation*}
    P(\mathbf{e}|\boldsymbol{\phi}^{(i)}) = \prod_{v=1}^{V}{(\mu_v^{(i,q_v)})^{e_v}(1-\mu_v^{(i,q_v)})^{1-e_v}}. 
\end{equation*}
To derive the new E and M steps, we follow Bishop \cite{bishop2006pattern}. We introduce an $\alpha$-dimensional latent variable $\mathbf{z} = (z_1,\ldots,z_\alpha)$ with binary elements and $\sum_{i=1}^\alpha z_i = 1$. Then the log-likelihood function of the complete data given the mixtures' parameters is:
\begin{multline*} \label{eq:syndromeguidedEM}
    \mathbb{E}\bigg[\log{P(\mathbf{e}^{(1)},\boldsymbol{q}^{(1)},\mathbf{z}^{(1)},\ldots,\mathbf{e}^{(K)},\boldsymbol{q}^{(K)},\mathbf{z}^{(K)}|\mathcal{R}^{\mathrm{syn}})}\bigg] = \\ =  \sum_{\kappa=1}^K  \sum_{i=1}^{\alpha}\mathrm{Res}_{\kappa,i} \bigg[\log\pi_i+ \\ + \sum_{v=1}^V  \Big(e_v^{(\kappa)}\log\mu_v^{(i,q_v^{(\kappa)})} + (1-e_v^{(\kappa)})\log(1-\mu_v^{(i,q_v^{(\kappa)})})\Big)\bigg].
\end{multline*}
The E-step becomes:
\begin{equation}
    \mathrm{Res}_{\kappa,i} = \frac{\pi_i  P(\mathbf{e}^{(\kappa)}|\boldsymbol{\phi}^{(i)})}{P(\mathbf{e}^{(\kappa)}|\mathcal{R}^{\mathrm{syn}})},
\end{equation}
where $\mathrm{Res}_{\kappa,i} \equiv \mathbb{E}[z_i^{(\kappa)}]$ is the responsibility of distribution $i$ given sample $\kappa$.

The new M-step is:
\begin{equation}
\label{eq:mu_syndrome_guided}
    \mu_v^{(i,b)} =
     \frac{\sum\limits_{\kappa=1}^K \mathbbm{1}_{q_v^{(\kappa)}=b}\mathrm{Res}_{\kappa,i} e_v^{(\kappa)}}{\sum\limits_{\kappa=1}^K \mathbbm{1}_{q_v^{(\kappa)}=b}\mathrm{Res}_{\kappa,i}}, \pi_i = \frac{\sum_{\kappa=1}^K \mathrm{Res}_{\kappa,i}}{K}
\end{equation}
with $b\in\{0,1\}$. In equation $(\ref{eq:mu_syndrome_guided})$, only the indices with active $q_v$ in $\boldsymbol{\mu}^{(i,q_v)}$ are updated with the new responsibilities. The data partition that follows this clustering is referred to as the syndrome-guided EM approach.

\subsection{Gating Function}
\label{sec:GatingFunction}

We consider three gating functions - \begin{enumerate*}[label=(\roman*)]
\item \textit{single-choice} gating,\label{item:single_choice}
\item \textit{all-decoders} gating and \label{item:all_decoders}
\item \textit{random-choice} gating.\label{item:random_gating}
\end{enumerate*}

\ref{item:single_choice} In reality, one does not have full knowledge of $\mathbf{e}$ at the decoder. To compensate for this knowledge-gap, we propose employing a classical non-learnable HDD between the channel's output and the ensemble. In this work, we chose to work with the Berlekamp-Massey algorithm \cite{berlekamp1968algebraic}. The HDD is employed to output estimated codeword $\tilde{\mathbf{c}}$, from which estimated error $\tilde{\mathbf{e}}=\mathbf{y}_{\mathrm{HD}}\xor \tilde{\mathbf{c}}$ is calculated. Then for each $\boldsymbol{\ell}$, we set $\mathcal{G}(\boldsymbol{\ell})_j=1$ for index $j$ realizing $\mathbf{\tilde{e}}\in\mathcal{X}^{(j)}$ and $\mathcal{G}(\boldsymbol{\ell})_i=0$ otherwise.

\ref{item:all_decoders} In comparison, the all-decoders gating function simply assigns $\mathcal{G}(\boldsymbol{\ell})_j=1$ for all $j$. The HDD remains unused. This gating serves as a baseline: the FER in the single-gating case is lower-bounded by the FER achievable by employing all decoders in an efficient manner.

\ref{item:random_gating} The second baseline adopted is the random-choice gating method, which assigns $\mathcal{G}(\boldsymbol{\ell})_j=1$ for a random $j$ and $\mathcal{G}(\boldsymbol{\ell})_i=0$ otherwise. Its target is to prove the significance of the single-choice gating.
\subsection{Combiner Mechanism}
\label{sec:Combiner}

The combination of decoded words is only considered in the case of the all-decoders gating:
\begin{prop}
    Take the score function $\mathcal{C}(\mathbf{\hat{c}}^{(i)})=\mathbf{\hat{c}}^{(i)}\boldsymbol{\ell}^\mathrm{T}$. Then the combination rule becomes: 
\begin{equation}
    \mathbf{\hat{c}} = {\argmin}_{\mathbf{\hat{c}^{(i)}}\in\{\mathbf{\hat{c}}^{(1)},\ldots,\mathbf{\hat{c}}^{(\alpha)}\}}\mathbf{\hat{c}}^{(i)}\boldsymbol{\ell}^\mathrm{T}
\end{equation}
\end{prop}

\begin{figure*}[!t]
\centering
\subfloat[CR-BCH(63,36)]{
    \includegraphics[width=0.47\textwidth,height=0.274\textheight]{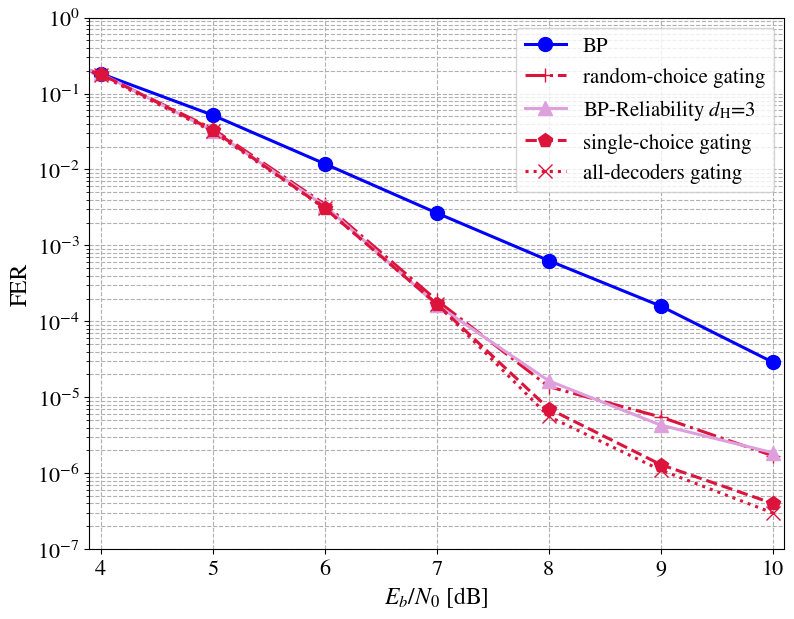}
    \includegraphics[width=0.47\textwidth,height=0.274\textheight]{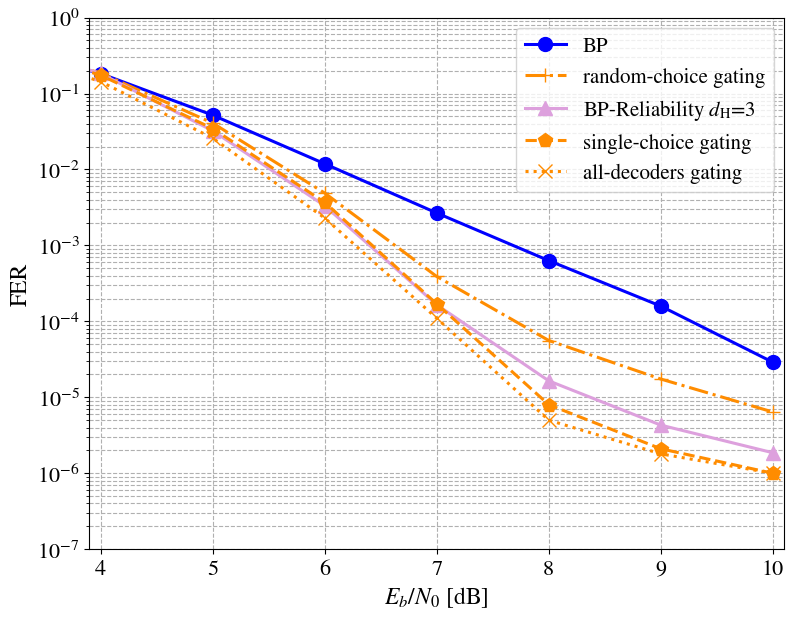}
    \label{subfig:63_36}
}

\subfloat[CR-BCH(63,45)]{
    \includegraphics[width=0.47\textwidth,height=0.274\textheight]{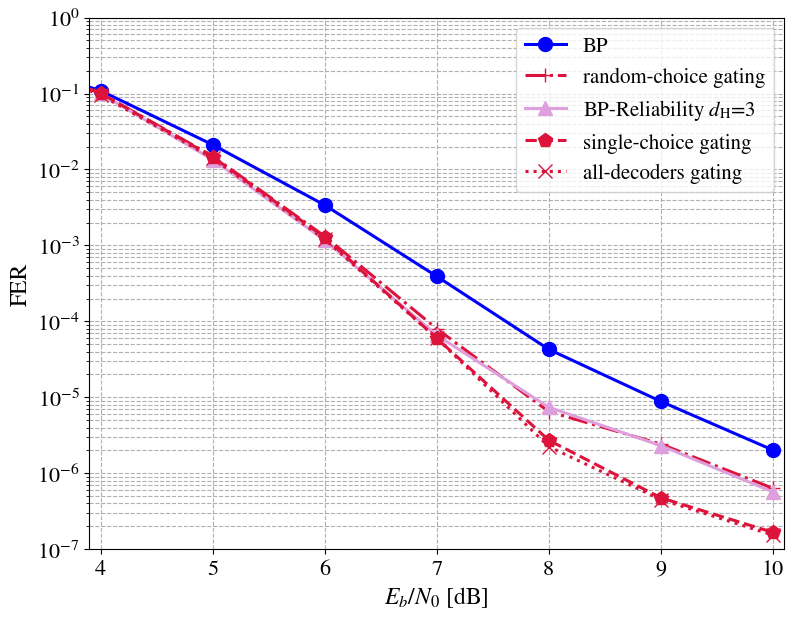}
    \includegraphics[width=0.47\textwidth,height=0.274\textheight]{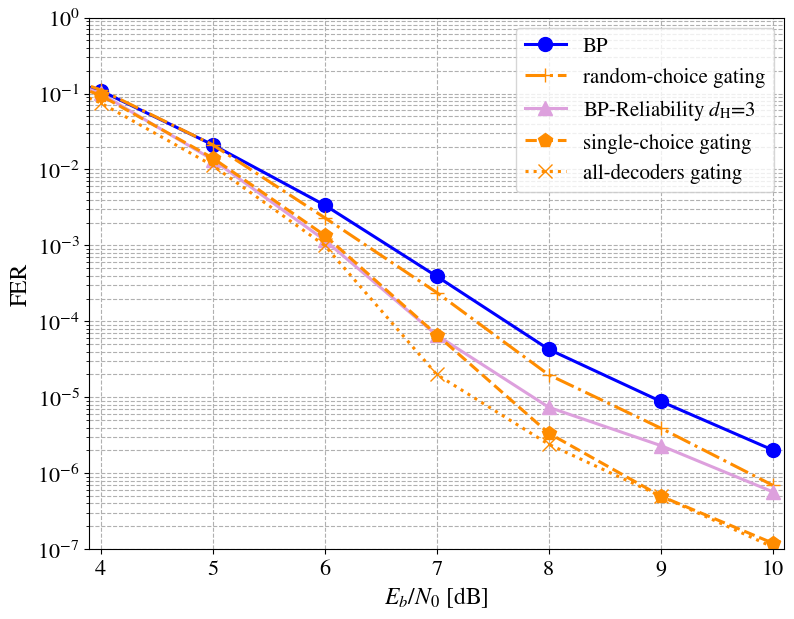}
    \label{subfig:63_45}
}
    
\caption{FER comparison: Hamming distance approach on the left side and syndrome-guided EM approach on the right side}
\label{fig:performance}
\end{figure*}

This particular score function has greater values for codewords than for pseudo-codewords (see proposition 2.12 in \cite{koetter2007characterizations}). Therefore, it mitigates the effects of the pseudo-codewords, which are most dominant at the error floor region as indicated in \cite{richardson2003error}. 

\renewcommand{\arraystretch}{1.1} 
\begin{table}[t!]
    \centering
      \begin{threeparttable}
        \caption{Training Hyperparameters}
          \label{tab:train}
            \begin{tabular}{| c  c |}
                \hline \textbf{Hyperparameters} & \textbf{Values}\\ 
                \hline
                Architecture & Feed Forward\\ 
                \hline 
                Initialization & as in [5] \\
                \hline
                Loss Function & Binary Cross Entropy with Multiloss 
                \\
                \hline
                Optimizer & RMSPROP \\
                \hline
                $\rho_t$ range & 4dB to 7dB 
                \\  
                \hline
                From-Scratch Learning Rate & 0.01 \\
                \hline
                Finetune Learning Rate & 0.001 
                \\
                \hline
                Batch Size & 1000 words per SNR 
                \\
                \hline
                Messages Range & $(-10,10)$ \\
                \hline
            \end{tabular}
      \end{threeparttable}
\end{table}

\section{Results}
\label{sec:Results}
We present results of simulating the ensembles based on the Hamming distance and the syndrome-guided EM approaches for two different linear codes BCH(63,45) and BCH(63,36). We use the cycle-reduced (CR) parity-check matrices as in \cite{helmling2016database}. Every WBP member is trained until convergence. The vectorized Berlekamp-Massey algorithm we have used is based on \cite{morelos2006art}. Training is done on zero codewords only, due to the symmetry of the BP (see \cite{nachmani2018deep} for further details). We consider 5-iterations only for BP decoding as the common benchmark. Syndrome based stopping criterion is applied after each BP-iteration. The validation dataset is composed of SNR values of 1dB to 10dB, at each point at least 100 errors are accumulated.

The number of decoders chosen was $\alpha=3$ for both methods, as adding decoders did not boost performance significantly. For the Hamming approach, the three regions chosen were $\mathcal{X}^{(1)},\mathcal{X}^{(2)},\mathcal{X}^{(3)}$. Training is done by finetuning, starting from weights of the BP-FF in \cite{nachmani2018deep}, with a smaller learning rate as specified in Table \ref{tab:train}. For the syndrome-guided EM approach, all decoders are trained from scratch, as finetuning yielded lesser gains. In the training phase, one assumes knowledge of the transmitted word. Thus, all datasets contained the known errors (no HDD employed in training). We empirically chose $K=10^6$, equally drawn from SNR values of 4dB to 7dB. These SNR values neither have too noisy words nor too many correct words \cite{gruber2017deep}. Relevant training hyperparameters appear in Table \ref{tab:train}. 

Figure \ref{fig:performance} presents the results of the simulation. One may observe that our proposed approaches compare favorably to the previous best results from \cite{be2019active} (the BP-Reliability approach) up to SNR of 7dB, and surpasses it thereafter. FER gains of up to 0.4dB at the waterfall region are observed for both approaches in the two codes. At the error floor region, the improvement varies from 0.5dB to 1.25dB in the CR-BCH(63,36), while a constant 1dB is observed in the CR-BCH(63,45). No improvement is achieved in the low-SNR regime. We attribute this to the limitation of the model-based approach, also seen in other model-based approaches \cite{nachmani2018deep,be2019active}.

The two methods have non-negligible performance difference only at SNR of 9dB and 10dB. The Hamming distance approach surpasses the syndrome-guided EM one in the CR-BCH(63,36) with the reverse situation in the CR-BCH(63,45). The gating for the Hamming approach is optimal, as indicated by the single-choice gating curve that adheres to the all-decoders lower-bound. The syndrome-guided gating is suboptimal over medium SNR values, as indicated by the gap between the single-choice gating and the all-decoders curves, having potential left for further investigation and exploitation. 

Lastly, comparing the random-choice gating for the two approaches, one can see that though the random-choice gating is worse for the syndrome-guided EM ensemble than for the Hamming distance ensemble, the gains of the two ensembles are quite similar. This hints that each expert in the EM method specializes on a smaller region of the input distribution, yet as a whole these experts complement one another, such that the syndrome-guided EM ensemble covers as much of the input distribution as the Hamming distance ensemble. This notion requires further exploration.

\section{Conclusions and Further Work}
This paper introduces a novel approach to machine-learning based channel decoding: the ensemble of decoders. This ensemble combines classical and trainable decoders in a hybrid manner. We have shown that remarkable gains may be achieved using this hybrid approach, therefore encouraging further research of this direction. Another open question is how to apply the framework to longer codes, while keeping the number of decoders moderate. As the complexity of the data will increase, more sophisticated partitions must be designed.
\section*{Acknowledgment}
We thank Ishay Be'ery for many insightful discussions.

\appendix
\section{Appendices}
\subsection{Proof Of Proposition 1}
\label{app:proof}

The proof is divided to two parts. First, we derive the \textit{true} centers of the mixture. Second, we show that the parameterized centers attain the global maximum of the likelihood function when identical to the true centers.

Concerning the first part, recall that the symmetric AWGN can be viewed as a binary symmetric channel (BSC) with crossover probability of $Q(\frac{1}{\sigma^{(i)}})$ per bit. In white noise settings, the i.i.d. claim holds, thus the vectorized crossover probabilities are $(Q(\frac{1}{\sigma^{(i)}}),\ldots,Q(\frac{1}{\sigma^{(i)}}))$ per word. Lastly, taking into account that the zero-errors vector is the one transmitted (as one always transmits a codeword), we can denote $\boldsymbol{\bar{\mu}}^{(i)}=(Q(\frac{1}{\sigma^{(i)}}),\ldots,Q(\frac{1}{\sigma^{(i)}}))$ and indeed every error drawn from channel $i$ is Bernoulli distributed with parameters $\boldsymbol{\bar{\mu}}^{(i)}$. This analysis is valid for each of the $\alpha$ channels separately, hence $\mathcal{\bar{R}}=\{(\boldsymbol{\bar{\mu}}^{(1)},\bar{\pi}_1),\ldots,(\boldsymbol{\bar{\mu}}^{(\alpha)},\bar{\pi}_\alpha)\}$ denotes the \textit{true} mixture (and see that $\bar{\pi}_i=\frac{1}{\alpha}$ for all values of $i$).      

Regarding the second part, due to the fact that the $log(\cdot)$ function is monotonic increasing, one can instead show that the global maximum is attained for the \textit{log-likelihood} function. Let us consider the log-likelihood function of the data samples given the parameters (equation (\ref{eq:log_likelihood})). At the infinite samples limit, the log-likelihood becomes:
\begin{equation*}
    \lim_{K\rightarrow \infty}\log{P(\mathcal{E}|\mathcal{R})} =  \mathop{\mathbb{E}}_{\mathbf{e}\sim\mathcal{\bar{R}}}\big[\log{P(\mathbf{e}|\mathcal{R})\big]}.
\end{equation*}
Now, plugging the entropy $\mathcal{H}$ and the Kullback–Leibler (KL) divergence denoted $KL$:
\begin{equation*}
    \mathop{\mathbb{E}}_{\mathbf{e}\sim\mathcal{\bar{R}}}\big[\log{P(\mathbf{e}|\mathcal{R})\big]} =  -\infdiv{P(\mathbf{e}|\mathcal{\bar{R}})}{P(\mathbf{e}|\mathcal{R})} - \mathcal{H}(P(\mathbf{e}|\mathcal{\bar{R}})).
\end{equation*}
As we are interested in parameters $\mathcal{R}$, the entropy term can be omitted:
\begin{multline*}
    \argmax_{\mathcal{R}}\mathop{\mathbb{E}}_{\mathbf{e}\sim\mathcal{\bar{R}}}\big[\log{P(\mathbf{e}|\mathcal{R})\big]} = \\ \argmin_{\mathcal{R}}\infdiv{P(\mathbf{e}|\mathcal{\bar{R}})}{P(\mathbf{e}|\mathcal{R})}.
\end{multline*}

At last, as the KL divergence is non-negative, the global maximum is attained when $\mathcal{R}$ contains the same centers as $\mathcal{\bar{R}}$. Put in different words, ${\boldsymbol{\mu}^{(i)}=\Big(Q(\frac{1}{\sigma^{(i)}}),\ldots,Q(\frac{1}{\sigma^{(i)}})\Big)}$, $\forall{i}$. This last part is motivated by the analysis in \cite{srebro2007there}.  

\bibliographystyle{IEEEtran}
\bibliography{references}

\end{document}